\documentclass{article}

\usepackage{arxiv}

\usepackage[utf8]{inputenc} 
\usepackage[T1]{fontenc}    
\usepackage{hyperref}       
\usepackage{url}            
\usepackage{booktabs}       
\usepackage{amsfonts}       
\usepackage{nicefrac}       
\usepackage{microtype}      
\usepackage{lipsum}		
\usepackage{framed,multirow}
\usepackage{stfloats}
\usepackage{xcolor}
\usepackage{amssymb}
\usepackage{latexsym}
\usepackage{graphicx}

\title{A Multi-View Dynamic Fusion Framework: How to Improve the Multimodal Brain Tumor Segmentation from Multi-Views?}

\author{
 Yi Ding\thanks{Yi Ding is with the Network and Data Security Key Laboratory of Sichuan Province, University of Electronic Science and Technology of China, Chengdu, Sichuan, 610054 China; he is also with Institute of Electronic and Information Engineering of UESTC in Guangdong, Guangdong, 523808, China (e-mail: yi.ding@uestc.edu.cn).} \\
  Department of Information \\
  and Software Engineering \\
  University of Electronic Science \\
  and Technology of China \\
  Chengdu, China \\
  \texttt{yi.ding@uestc.edu.cn} \\
   \And
 Wei Zheng \\
  Department of Information \\
  and Software Engineering \\
  University of Electronic Science \\
  and Technology of China \\
  Chengdu, China \\
   \And
 Guozheng Wu \\
  National Natural Science  \\
  Foundation of China \\
     \And
 Ji Geng \\
  Department of Information \\
  and Software Engineering \\
  University of Electronic Science \\
  and Technology of China \\
  Chengdu, China \\
     \And
 Mingsheng Cao \\
  Department of Information \\
  and Software Engineering \\
  University of Electronic Science \\
  and Technology of China \\
  Chengdu, China \\
     \And
 Zhiguang Qin \\
  Department of Information \\
  and Software Engineering \\
  University of Electronic Science \\
  and Technology of China \\
  Chengdu, China
}

\begin{document}
\maketitle

\begin{abstract}
When diagnosing the brain tumor, doctors usually make a diagnosis by observing multimodal brain images from the axial view, the coronal view and the sagittal view, respectively. And then they make a comprehensive decision to confirm the brain tumor based on the information obtained from multi-views. Inspired by this diagnosing process and in order to further utilize the 3D information hidden in the dataset, this paper proposes a multi-view dynamic fusion framework to improve the performance of brain tumor segmentation. The proposed framework consists of 1) a multi-view deep neural network architecture, which represents multi learning networks for segmenting the brain tumor from different views and each deep neural network corresponds to multi-modal brain images from one single view and 2) the dynamic decision fusion method, which is mainly used to fuse segmentation results from multi-views as an integrate one and two different fusion methods, the voting method and the weighted averaging method, have been adopted to evaluate the fusing process. Moreover, the multi-view fusion loss, which consists of the segmentation loss, the transition loss and the decision loss, is proposed to facilitate the training process of multi-view learning networks so as to keep the consistency of appearance and space, not only in the process of fusing segmentation results, but also in the process of training the learning network. \par

By evaluating the proposed framework on BRATS 2015 and BRATS 2018, it can be found that the fusion results from multi-views achieve a better performance than the segmentation result from the single view and the effectiveness of proposed multi-view fusion loss has also been proved. Moreover, the proposed framework achieves a better segmentation performance and a higher efficiency compared to other counterpart methods.
\end{abstract}


\section{Introduction}
\label{sec1}
An accurate and reliable brain tumor segmentation plays a key role in the diagnosis and treatment for brain tumors. However, the brain tumor segmentation based on MRI data is still a challenging task, including some key problems such as: 
\begin{enumerate}
\item Brain glioma can appear anywhere in the brain and it is varied in shape, appearance, size and type.
\item Brain glioma will invade the surrounding brain tissue instead of shifting, causing its boundary to be blurred\cite{ref10}.
\end{enumerate} \par
\begin{figure}[!ht]
\centering
\includegraphics[width=0.45\textwidth]{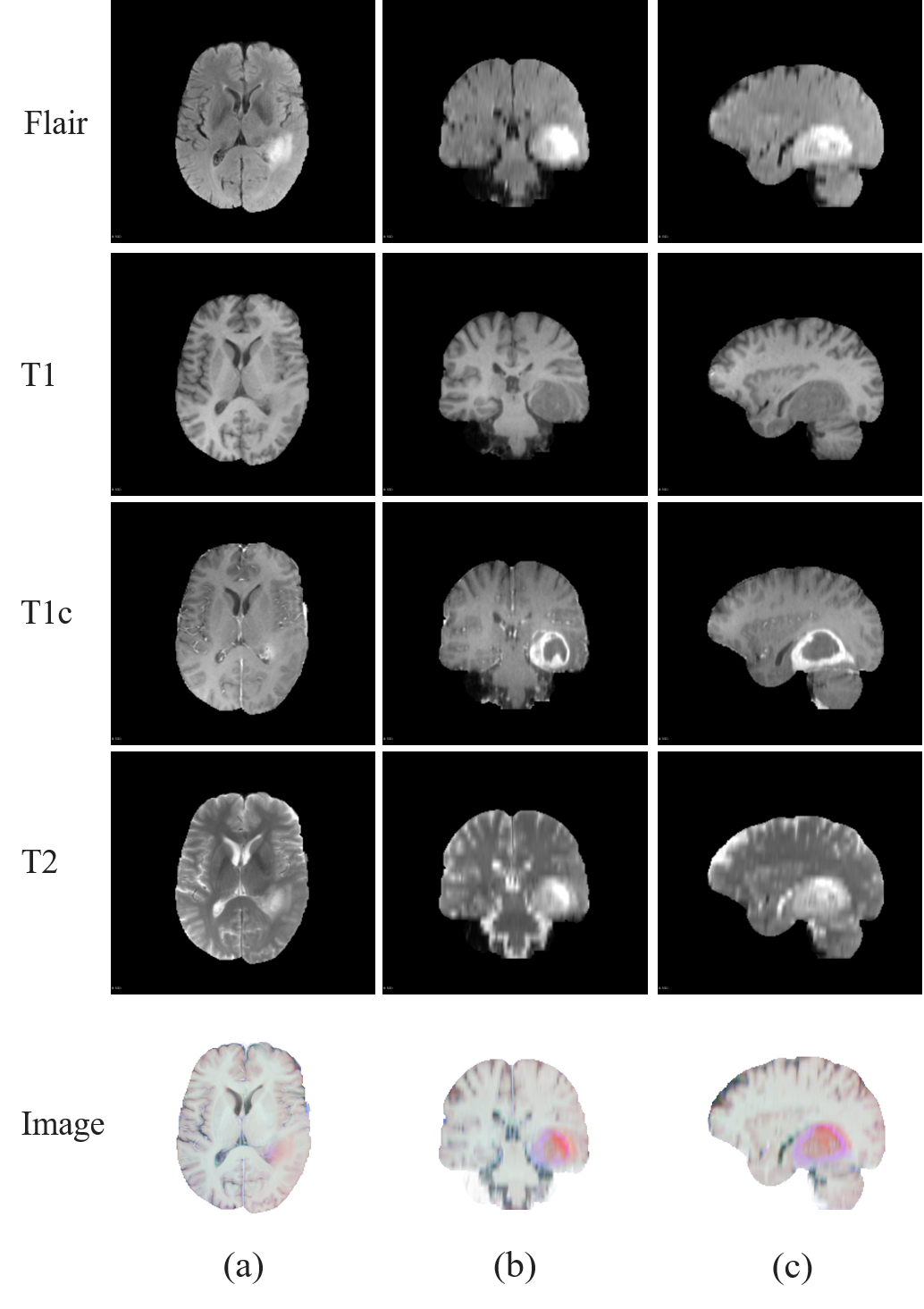}
\caption{Observing the brain tumor from different views, the FLAIR, T1, T1c, T2 indicate four modalities of MRI images. (a): The image in Axial View, (b): The image in Coronal View, (c): The image in Sagittal View.}
\label{fig:Multi-view_brain_tumor}
\end{figure}

Many researches have been developed to segment the brain tumor in an automatic way and the deep learning-based method is one of the most popular. These methods achieve a good performance for brain tumor segmentation. When processing with the magnetic resonance imaging (MRI) dataset which is actually 3D, most deep learning algorithms split the 3D brain dataset into several 2D slices and then train the deep learning network with these 2D slices. Furthermore, these deep learning-based segmentation methods usually segment the brain tumor by only adopting 2D slices from the axial view. However, in the clinical practice, the doctor makes a diagnosis by observing the brain images from the axial view, the coronal view and the sagittal view, respectively and then make a comprehensive decision based on the information obtained from multi-views. Different views can provide the useful information from different aspects and produces a great influence on each other. As shown in Figure \ref{fig:Multi-view_brain_tumor}, the image from the axial view contains more feature information about the brain area, but we can't see the lesion areas from this view. These lesion areas usually include the detailed information about the tumor area interested by doctors. In order to obtain this information, the doctor needs to further observe brain images from the coronal view and sagittal view, which intuitively include the position and shape information of brain tumor. Combining these three most common views, the doctor has the ability to make a comprehensive diagnosis and treatment for brain tumor so as to achieve a better clinical result. This diagnosis method also follows the principle of how human being observation objects in the real-world environment. We usually observe 3D objects from multi-views in the three-dimensional space. The object information from different views will be continuously fed back to the brain from eyes, and the brain conducts an integrated dynamic decision process to achieve a better human cognition on these objects. \par

Instead of segmenting the brain tumor only from 2D slices in single view, many efforts have been made to consider how to utilize the 3D brain information. The 3D-CNN holds a great potential to make full use of the 3D information of MRI data, which also contains multi-views information. But the 3D-CNNs-based brain tumor segmentation greatly increases the network scale and computational cost \cite{ref28,ref29}. Moreover, the 3D-CNN does not follow the way that doctors observe the brain images from three different views to make a diagnosis in clinical practice. Another way is to split the 3D dataset into 2D slices with the axial view, coronal view and sagittal view, respectively. And then different 2D-CNN networks are trained with these slices from multi-views to segment the brain tumor, such as in the \cite{ref16} and \cite{ref17}. The advantage is that the 2D-CNN based network can greatly reduce the computing cost and achieve a higher efficiency. But when fusing network results from multi-views into an integrate brain tumor segmentation, the fusing processing only considers the appearance and space consistency on the final segmentation results, while the appearance and space consistency should also be kept during the network training. \par

Inspired by the above works, this paper proposes a multi-view dynamic fusion framework to improve the performance of brain tumor segmentation. The novel idea is based on two important insights: (1) The learning process can be designed by following the way that doctors observe the MRI images from the axial view, coronal view, and sagittal view, respectively and then make an integrated decision according to the tumor information obtained from multi-views; (2) In order to further make use of the 3D information, the appearance and space consistency of brain tumor on multi-views can be regarded as a factor to guide both the network training process and the segmentation result fusing process to achieve a better performance. The proposed framework consists of two parts: (1) a multi-view deep neural network architecture; (2) a dynamic decision fusion method. The former one is mainly used to obtain the brain tumor segmentation results from different views. Each learning network corresponds to 2D images from one single view. After obtaining different segmentation results, the dynamic decision fusion method is adopted to integrate these results as a final one. In this paper, the voting method and the weighted averaging method have been adopted as the fusion method to carry on the mutual comparative analysis. More importantly, the multi-view fusion loss, which consists of the segmentation loss, the transition loss and the decision loss, is proposed to ensure the consistency of appearance and space in the process of training the learning network. Furthermore, some simple yet effective improvements, which include the supplemental input and the post-processing step, have also been proposed to improve the brain tumor segmentation. \par

In a nutshell, the main contributions of this paper are summarized as follows:
\begin{enumerate}
\item A multi-view dynamic fusion framework is developed to realize the brain tumor segmentation by following the process of making the diagnosis in clinical practice.
\item The multi-view fusion loss is proposed to ensure the appearance and spatial consistency so as to improve the network training process. 
\item Extensive experiments are conducted on the BRATS 2015 and 2018 datasets to evaluate the proposed framework. The results demonstrate that the better segmentation performance can be obtained by fusing results from multi-views. Moreover, compared with existing brain tumor segmentation method, the proposed method can achieve a better performance both on effectiveness and efficiency.
\end{enumerate} \par

The remainder of this paper is as follows: Section \ref{section:Related Work} gives an introduction of the deep learning and brain tumor segmentation methods. Section \ref{section:The Multi-View Dynamic Fusion Framework} presents the details of the proposed multi-view dynamic fusion framework. Section \ref{section:Experiments, Results, and Discussion} shows the segmentation performance and evaluates the network efficiency. Section \ref{section:Conclusion} gives a summarization. 

\section{Related Work} \label{section:Related Work}

\subsection{Deep Neural Network}
In recent years, with the continuous improvement of computing power \cite{ref18}, the deep neural network has been revived and become a powerful supervised learning tool with efficient data modeling capabilities and the ability to highly distinguish specific task data features \cite{ref2,ref1}. The deep learning method has made significant progress in the fields of image classification and image segmentation \cite{ref2,ref1,ref3}, such as the superior performance of various deep neural network models \cite{ref2,ref4,ref5} in various natural image datasets such as the ImageNet, coco and City-scapes. From the earliest Alexnet \cite{ref1} to FCN \cite{ref19}, VGG \cite{ref20} and other network models have achieved good results on ImageNet. Compared with the FCN model, the UNET model \cite{ref21} has achieved better results in biomedical image segmentation during the same period. After that, deep neural network model is also being continuously updated, and high-performance neural network models such as GoogleNet \cite{ref22}, ResNet \cite{ref2}, DesNet \cite{ref4}, SE-Net \cite{ref23} and recent EfficienNet \cite{ref5} have been proposed in succession, these models consider not only the extension of the network model but also how to extract the high-level features corresponding to the data more efficiently. The EfficienNet-B7 network model \cite{ref5} has increased the accuracy of the top-1 index and top-5 index on ImageNet to 84.4\% and 97.1\%, respectively, which has greatly exceeded human performance in this dataset. \par
However, most of these current deep learning methods are based on 2D natural images. Although there are some 3d image processing methods \cite{ref6,ref7}, most of them focus on the time series of images, which means the space-time dimension of images. These methods pay less attention to the difference in the multi-views of 3D images, which means the appearance and space consistency of images.

\begin{figure*}[htbp]
\centering
\includegraphics[width=\textwidth]{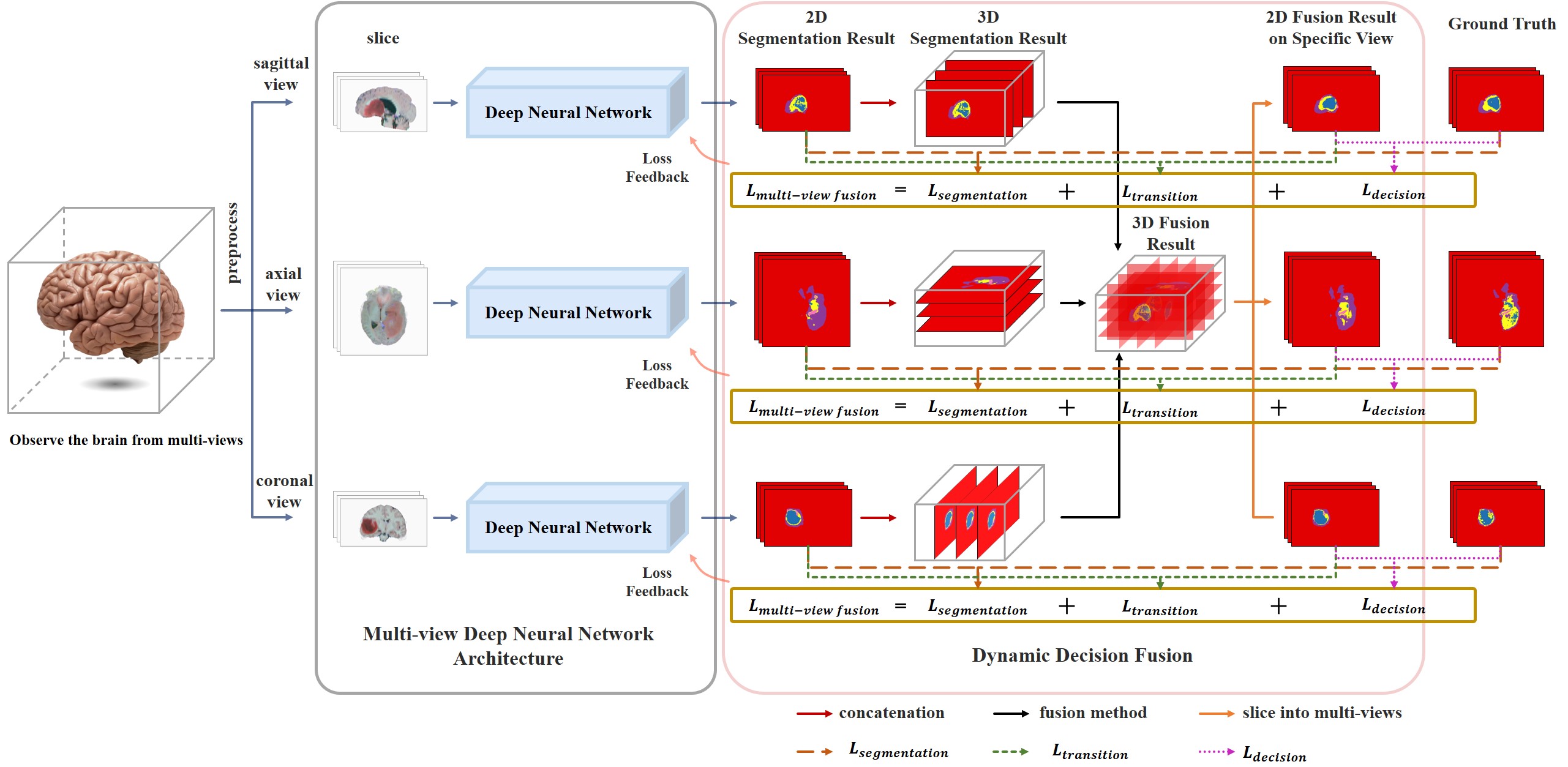}
\caption{The overall architecture of the Multi-View Dynamic Fusion Framework for segmenting the 3D brain tumor from the axial view, coronal view and sagittal view.}
\label{fig:Overall_Architecture}
\end{figure*}

\subsection{Medical Image Segmentation}
In recent years, with the continuous development of deep learning, researchers were also expanding the application of deep learning technology in the field of medical imaging \cite{ref43}. \cite{ref24} first proposed the use of GPU to implement 2D CNN for neuronal cell membrane segmentation; \cite{ref25}, \cite{ref26}, \cite{ref14} and \cite{ref41} proposed a variety of brain tumor segmentation methods based on deep learning models in BRATS 2015 open challenge \cite{ref27}. They used different deep learning models, including CNNs, convolutional restricted Boltzmann machines, and superimposed denoising autoencoder, though their model structure is simple, these early methods have achieved certain effects, which show the potential of deep learning. \par 
In the methods of brain tumor image segmentation based on deep learning, methods based on CNNs have achieved better segmentation effects, especially those using 2D-CNNs or 3D-CNNs models to construct the segmentation model. Although 3D-CNNs has the potential to make full use of the 3D information of MRI data, it also greatly increases the network scale and computational cost. \cite{ref28} and \cite{ref29} pointed out the problems and limitations of using 3D CNN on medical imaging data. Therefore, 2D-CNNs has been widely used in the segmentation of brain tumors: \cite{ref26} used very deep CNNs to segment the brain tumor by stages; \cite{ref30} modeled multiple brain tumor segmentation tasks into three subtasks and each subtask was solved by CNNs; \cite{ref14} proposed a deep learning method with two CNNs pathways, including convolution pathway and fully connected pathway; \cite{ref39} designed the deep CNNs to segment the brain tumor by considering the symmetry of the brain. \par
These brain tumor segmentation methods classify each brain image into different classes, including healthy tissue, necrosis, edema, non-enhancing tumors, and enhancing tumors, and then mark the classification results of each image with central voxels to achieve brain tumor segmentation. The above-mentioned brain tumor segmentation methods based on CNNs-based mostly assume that the labels of each voxel are independent, without considering the appearance and space consistency of brain tumors \cite{ref17}. In order to consider the appearance dependence of the label, \cite{ref31} and \cite{ref14} constructed the cascaded network model structure by taking the pixel-level probability segmentation results obtained by the early training CNNs as additional input of subsequent CNNs, on the other hand, \cite{ref32} combining Markov Random Fields (MRFs) especially conditional random fields (CRFs) and deep learning techniques for brain tumor image segmentation studies. It can be used as a post-processing step for CNNs to achieve better segmentation performance \cite{ref33,ref34}, which also applies the idea of appearance and space consistency. \par

Multiple 2D CNNs can be integrated to segment 3D medical images: a special three-plane CNNs for knee-joint cartilage segmentation is proposed by the \cite{ref28}, which used three neural networks to process image patches extracted from xy, yz and zx planes, and then the three plane outputs are fused by the softmax classifier. \cite{ref35} proposed a pseudo-3D patch-based approach in which the image patches in the axial, coronal and sagittal view are separately convoluted, and then merged in the fully connected layer. \cite{ref36} used multi-view streams to detect brain tumors. They proposed a network structure consisting of 2D CNNs in axial, coronal and sagittal. Each of them was used to process patches extracted from the specific view of brain tumors. The output of multi-view 2D CNNs streams is then fused to segment the brain tumor image; \cite{ref17} combined the methods of 3D CRF with multi-view streams and proposed a set of proprietary post-processing methods to segment brain tumor images. However, most of these methods are based on a specific neural network model, which uses image patches for network training. They only adopt one kind of decision fusion method to consider the appearance and space consistency of the results from multi-views, and they do not consider the appearance and space consistency in the process of neural network training.

\section{The Multi-View Dynamic Fusion Framework} \label{section:The Multi-View Dynamic Fusion Framework}

\begin{figure*}[!ht]
\centering
\includegraphics[width=\textwidth]{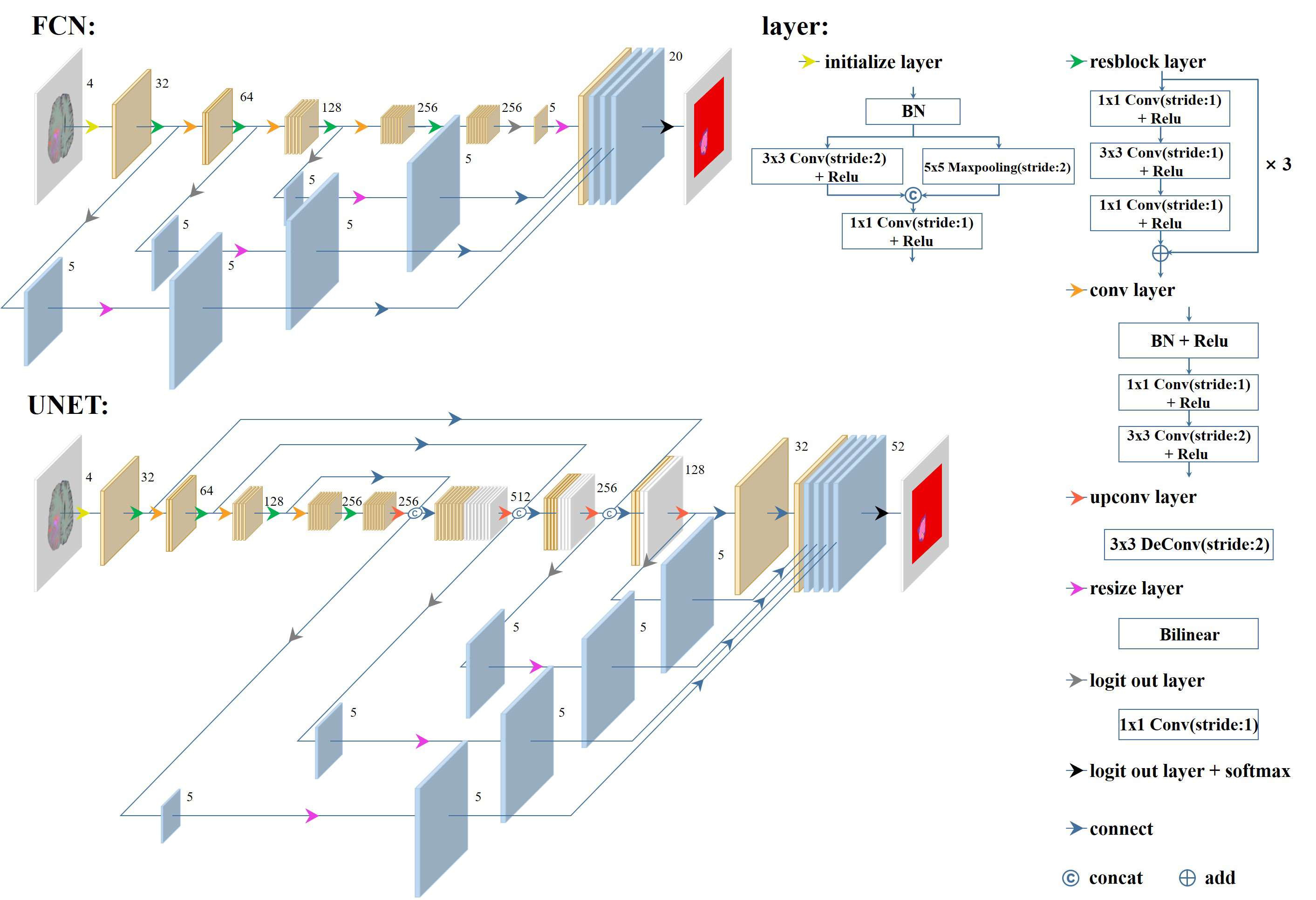}
\caption{The FCN network structure and UNET network structure for the multi-view network. The network operation layer includes the initialization layer, resblock layer, conv layer, upconv layer, resize layer, logit out layer and softmax, where BN represents the batch normalization layer, Relu represents the activation function, Conv represents the downsampling operation, DeConv represents the upsampling operation, and Bilinear represents the bilinear interpolation.}
\label{fig:Network_Architecture}
\end{figure*}

\subsection{Overall Architecture}
As shown in Figure \ref{fig:Overall_Architecture}, the overall architecture of proposed framework mainly consists of the multi-view deep neural network architecture and the dynamic decision fusion. The former one is mainly used to segment the brain tumor from the axial view, the coronal view and the sagittal view respectively, while the dynamic decision fusion module is designed to dynamically fuse these segmentation results as an integrate one by utilizing more useful information from multi-views so as to achieve a better segmentation performance. When adopting this framework to segment the brain tumor, the input is the 3D brain dataset of one patient at each time. The 3D dataset is normalized to follow the standard normal distribution as the preprocessing step. Then, the normalized 3D data is sliced into 2D images from the axial view, the coronal view and the sagittal view, and are regarded as the input of multi-view networks. Each branch learning network is adopted as the segmentation network for processing 2D slices from one single view. After obtaining the segmentation result of each 2D slice from multi-views, these 2D results are concatenated together and reconstructed as the 3D tumor segmentation result from the axial view, the coronal view and the sagittal view respectively. The dynamic decision fusion method is then employed to make a comprehensive processing of these 3D tumor segmentations from multi-views, presenting the 3D fusion segmentation result. Moreover, in order to facilitate the process of multi-view network training, the proposed multi-view fusion loss is employed to calculate the segmentation loss (the loss between the 2D segmentation result and the ground truth), the decision loss (the loss between the fusion result on specific view and the ground truth) and the transition loss (the loss between 2D segmentation result and the fusion result in specific view). Based on the multi-view fusion loss, the back-propagation algorithm can be used to dynamically update the training process of multi-view network to achieve a better 2D segmentation performance. 

\subsection{Multi-View Deep Neural Network Architecture} \label{section:Network Architecture}

The multi-view network consists of multiple feature extraction networks in a parallel way, where each network is trained to extract the useful features from brain tumor images in one specific view. In this work, two different network architectures are evaluated for the multi-view network. One is the Fully Convolutional Network (FCN) structure, which is composed of the downsampling operation and the bilinear interpolation operation. While another one is the UNET network structure, which mainly consists of the downsampling operation and the upsampling operation. The Figure \ref{fig:Network_Architecture} presents the network structure of the FCN and the UNET. The down-sampling process of these two networks is kept the same. 

For each feature extraction network, the multimodal MRI images in a form of 3D are firstly sliced into 2D images from a specific view. And these 2D slices from the T1, T1c, T2 and Flair models are concatenated together to obtain the original input image with a size of $H*W*4$, where H and W represent the height and width of the image. Then, the original input images are processed by an initialization layer to extract initial feature maps with a size of $H/2*W/2*32$. After the resblock layer is processed (except for the last resblock layer), the number of channels is doubled, and the image size is reduced by half after each conv layer. The network structure of the downsampling process is shown in Table \ref{tab:structure}. In the FCN network, after the logit out layer is processed, the number of feature map channels in each stage of the downsampling process is equal to the number of tumor classes. For example, the brain tumor is classified into 5 categories, therefore the number of the feature map channel in the first stage of the downsampling process is set to 5 after the logit out layer. These outputs are then restored to the original image size of $H*W$ by employing the bilinear interpolation method. Finally, the pixel-by-pixel class probability can be calculated based on the logit out layer and the softmax function.

\begin{table}[!htbp]
  \centering
  \caption{The structure of the deep neural network in the downsampling stage.}
  \resizebox{0.49 \textwidth}{!}{ 
    \begin{tabular}{c|c|cc}
    \hline
    layer name & output size & \multicolumn{2}{c}{convolution} \\
    \hline
    input & $H$ $\times$ $W$ & \multicolumn{2}{c}{——} \\
    \hline
    \multirow{2}[1]{*}{initialize} & \multirow{2}[1]{*}{
    ${H \mathord{\left/{\vphantom {H 2}} \right. \kern-\nulldelimiterspace} 2}$  $\times$ ${W \mathord{\left/{\vphantom {H 2}} \right. \kern-\nulldelimiterspace} 2}$} & \multicolumn{1}{c|}{3 $\times$ 3, 32, stride 2 } & 5 $\times$ 5 max pool, stride 2 \\
\cline{3-4}          &       & \multicolumn{2}{c}{1 $\times$ 1, 32, stride 1} \\
    \hline
    \rule{0pt}{28pt}
    resblock 1 & ${H \mathord{\left/{\vphantom {H 2}} \right. \kern-\nulldelimiterspace} 2}$ $\times$ ${W \mathord{\left/{\vphantom {H 2}} \right. \kern-\nulldelimiterspace} 2}$ & 
  \multicolumn{2}{c}{ $\left[ {{\rm{ }}\begin{array}{*{20}{c}}
{{\rm{1 }} \times {\rm{ 1, 64}}}\\
{{\rm{3 }} \times {\rm{ 3, 64}}}\\
{{\rm{1 }} \times {\rm{ 1, 64}}}
\end{array}{\rm{, stride\ 1}}} \right]{\rm{ }} \times {\rm{3}}$ } \\[18pt]
    \hline
    conv 1 & ${H \mathord{\left/{\vphantom {H 2}} \right. \kern-\nulldelimiterspace} 4}$ $\times$ ${W \mathord{\left/{\vphantom {H 2}} \right. \kern-\nulldelimiterspace} 4}$ & \multicolumn{2}{c}{ $ \begin{array}{*{20}{c}}
{{\rm{1 }} \times {\rm{ 1, 64, stride\ 1}}}\\
{{\rm{3 }} \times {\rm{ 3, 64, stride\ 2}}}
\end{array} $ } \\
    \hline
    \rule{0pt}{28pt}
    resblock 2 & ${H \mathord{\left/{\vphantom {H 2}} \right. \kern-\nulldelimiterspace} 4}$ $\times$ ${W \mathord{\left/{\vphantom {H 2}} \right. \kern-\nulldelimiterspace} 4}$ & \multicolumn{2}{c}{ $\left[ {{\rm{ }}\begin{array}{*{20}{c}}
{{\rm{1 }} \times {\rm{ 1, 64}}}\\
{{\rm{3 }} \times {\rm{ 3, 64}}}\\
{{\rm{1 }} \times {\rm{ 1, 128}}}
\end{array}{\rm{, stride\ 1}}} \right]{\rm{ }} \times {\rm{3}}$ } \\[18pt]
    \hline
    conv 2 & ${H \mathord{\left/{\vphantom {H 2}} \right. \kern-\nulldelimiterspace} 8}$ $\times$ ${W \mathord{\left/{\vphantom {H 2}} \right. \kern-\nulldelimiterspace} 8}$ & \multicolumn{2}{c}{ $ \begin{array}{*{20}{c}}
{{\rm{1 }} \times {\rm{ 1, 128, stride\ 1}}}\\
{{\rm{3 }} \times {\rm{ 3, 128, stride\ 2}}}
\end{array} $ } \\
    \hline
    \rule{0pt}{28pt}
    resblock 3 & ${H \mathord{\left/{\vphantom {H 2}} \right. \kern-\nulldelimiterspace} 8}$ $\times$ ${W \mathord{\left/{\vphantom {H 2}} \right. \kern-\nulldelimiterspace} 8}$ & \multicolumn{2}{c}{ $\left[ {{\rm{ }}\begin{array}{*{20}{c}}
{{\rm{1 }} \times {\rm{ 1, 64}}}\\
{{\rm{3 }} \times {\rm{ 3, 64}}}\\
{{\rm{1 }} \times {\rm{ 1, 256}}}
\end{array}{\rm{, stride\ 1}}} \right]{\rm{ }} \times {\rm{3}}$ } \\[18pt]
    \hline
    conv 3 & ${H \mathord{\left/{\vphantom {H 2}} \right. \kern-\nulldelimiterspace} 16}$ $\times$ ${W \mathord{\left/{\vphantom {H 2}} \right. \kern-\nulldelimiterspace} 16}$ & \multicolumn{2}{c}{ $ \begin{array}{*{20}{c}}
{{\rm{1 }} \times {\rm{ 1, 256, stride\ 1}}}\\
{{\rm{3 }} \times {\rm{ 3, 256, stride\ 2}}}
\end{array} $ } \\
    \hline
    \rule{0pt}{28pt}
    resblock 4 & ${H \mathord{\left/{\vphantom {H 2}} \right. \kern-\nulldelimiterspace} 16}$ $\times$ ${W \mathord{\left/{\vphantom {H 2}} \right. \kern-\nulldelimiterspace} 16}$ & \multicolumn{2}{c}{ $\left[ {{\rm{ }}\begin{array}{*{20}{c}}
{{\rm{1 }} \times {\rm{ 1, 64}}}\\
{{\rm{3 }} \times {\rm{ 3, 64}}}\\
{{\rm{1 }} \times {\rm{ 1, 256}}}
\end{array}{\rm{, stride\ 1}}} \right]{\rm{ }} \times {\rm{3}}$ } \\[18pt]
    \hline
    \end{tabular}
    }
  \label{tab:structure}
\end{table}

Different from the FCN network, the UNET network doesn't directly adopt the bilinear interpolation operation for recovering the image. Instead, the size of feature map is doubled after each upconv layer, and the number of feature map channels is reduced by half after each upsampling layer (except for the first stage). The upsampling feature map is concatenated with the feature map which has the same size in the downsampling stage. And combined feature maps are set as the final output of the upconv layer. Another difference between these two networks is which feature maps are chosen to compose the final network output. As shown in Figure \ref{fig:Network_Architecture}, for the FCN network, the feature map from each stage in the downsampling process is resized and then concatenated together as the final output while the UNET network chooses the feature map from each stage in the upsampling process.

\subsection{Dynamic Decision Fusion Method} \label{section:Fusion Method}

After obtaining the segmentation results from multi-views, the next step is to fuse these results as an integrate one. In this paper, two different data decision fusion methods have been adopted to realize the fusion process, which are the voting method and the weighted averaging method.

\subsubsection{The Voting Method}
\begin{figure*}[ht]
\centering
\includegraphics[width=0.95\textwidth]{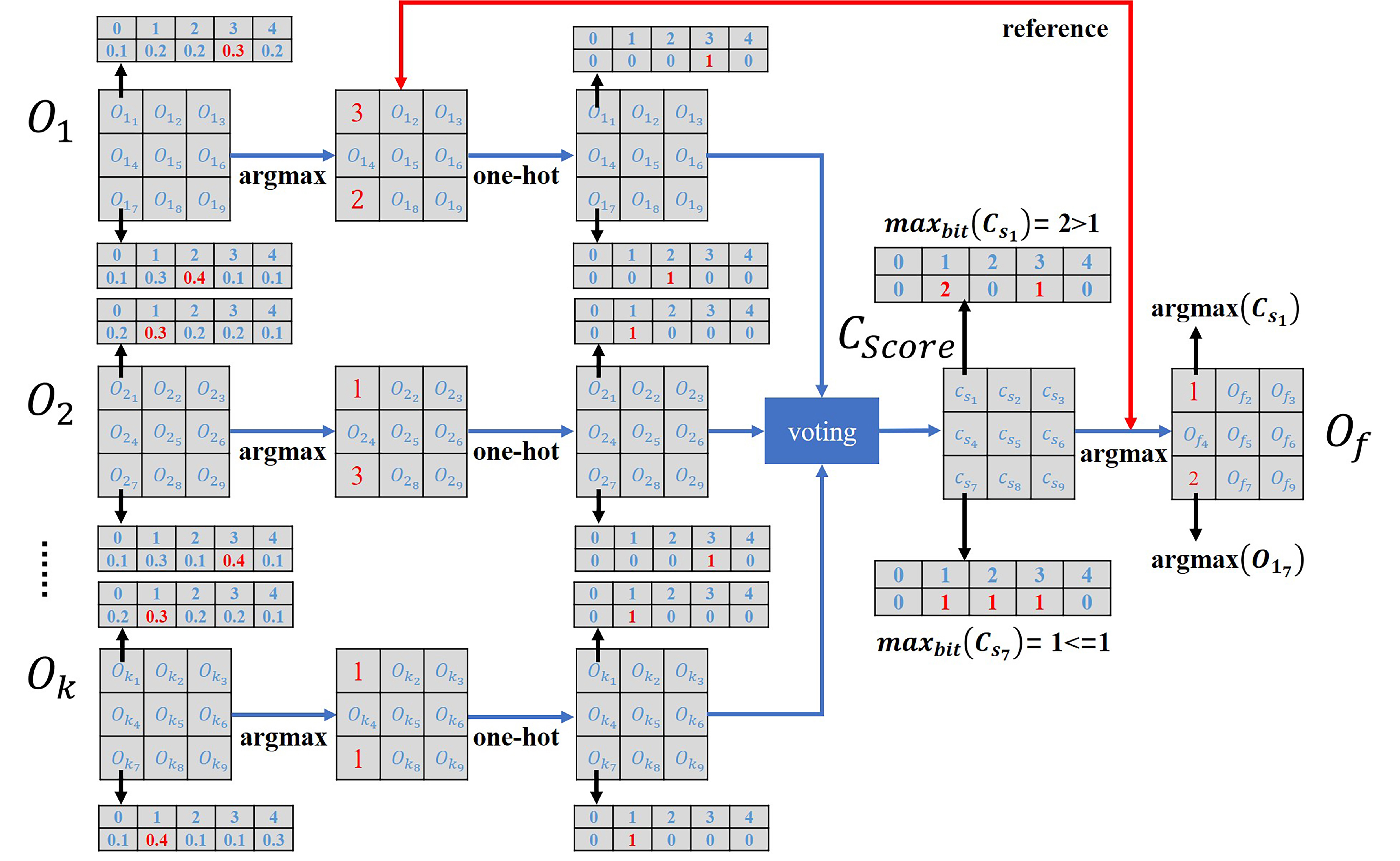}
\caption{An example of adopting the voting method with reference to the view 1 to perform the data fusion.}
\label{fig:Voting}
\end{figure*}

The voting method is a non-linear operation process, and an example of the voting method is shown in Figure \ref{fig:Voting}. The ${O_1}$, ${O_2}$ and ${O_3}$ represent the segmentation results from the axial, coronal and sagittal view respectively. The ${O_{1_n}}$, ${O_{2_n}}$ and ${O_{k_n}}$ in the matrix separately indicate a voxel with the same spatial position in different 3D segmentation results. Each voxel is labeled with the probability of different classes which it belongs to. There are five classes for labeling the brain tumor: 1 for necrosis; 2 for edema; 3 for non-enhancing tumor; 4 for enhancing tumor; 0 for everything else. For example, the ${O_{1_1}}$ is classified as the probability of 0.1, 0.2, 0.2, 0.3 and 0.2 corresponding to the class 0, 1, 2, 3 and 4, respectively. The voting method firstly adopts the argmax function, which is used to achieve the highest probability, to determine the class of voxel. In Figure \ref{fig:Voting}, the ${O_{1_1}}$ is set as class 3 and ${O_{2_1}}$ is set as class 1. Then, the one-hot encoding method is employed to calculate the probability of voxel again. The one-hot encoding translates the probability of voxel as a binary representation with only one significant bit. It means that the probability of other classes are set as 0 while the significant class is set as 1. For the ${O_{1_1}}$, the probability can be presented as 00010, which means that this voxel is classified as the class 3. And the ${O_{2_1}}$ is expressed as 01000. After obtaining the binary representation for each voxel in multi-views, the voting method fuses the voxel in the same spatial position from multi-views as an integrate one by re-calculating the probability of each voxel. In this paper, the ${C_{S_i}}$ is used to indicate the probability of each voxel in fused segmentation results and this score is calculated based on the following formula:

\begin{equation}
{C_{S_i}} = \sum\nolimits_{i = 1}^k {{f_{one\_hot}}(argmax({O_{i_j}}))} 
\end{equation}

Where $ k \ge 1$ and $ k \le 3$, argmax() function is used to determine the class of voxel in the original segmentation result while the one-hot encoding translates the class probability of voxel as the binary representation. The formula means that the updated probability for each voxel in fused segmentation results is calculated by adding the corresponding bit in the binary representation of each voxel with the same spatial position from multi-views. For example, the probability of ${C_{S_1}}$ is the sum of 00010 for ${O_{1_1}}$, 01000 for ${O_{2_1}}$, and 01000 for ${O_{k_1}}$, which is 02010. And the binary representation of ${C_{S_7}}$ is expressed as 01110. The last step of voting method is to confirm the class for each voxel in the fused one. The argmax function is adopted again to determine the voxel belongs to which class. This determination process is calculated by the following formula: 

\begin{equation}
{O_{f_i}} = \left\{ {\begin{array}{l} 
{argmax({C_{S_i}}),{\rm{if}}\ {max_{bit}}({C_{S_i}}) > 1}\\
{argmax({O_{ref}}),{\rm{if}}\ {max_{bit}}({C_{S_i}}) <= 1}\\
\end{array}} \right.
\end{equation}

where ${O_{f_i}}$ represents the final decision for each voxel in the fused segmentation result. The ${max_{bit}}$ is used to obtain the bit with the largest value in the binary representation, such as the ${max_{bit}(02010) = 2}$. If the value of ${max_{bit}}$, also called as the voting score of the voxel, is greater than 1, it indicates that most segmentation results from multi-views classify this voxel as the same class and the class with the highest score is regarded as the final decision for the current voxel. If the value of ${max_{bit}}$ is less than or equal to 1, it means that different segmentation results from multi-views fail to reach an agreement and this voxel can belong to any class. In this way, directly employing the voting result may lead to the wrong choice. Therefore, instead of adopting the voting result, the class with the highest probability of corresponding voxel in the reference view, represented as ${O_{ref}}$, is regarded as the final choice for the current voxel. For example, as shown in Figure \ref{fig:Voting}, for the ${C_{S_1}}$, the ${max_{bit}(C_{S_1}) = 2 > 1}$. Therefore, the highest probability in the voting result is adopted as the final choice for this voxel. For the ${C_{S_1}}$, the result of $argmax({C_{S_1}})$ is the final choice for this voxel, which is the class 1. For the ${C_{S_7}}$, the ${max_{bit}(C_{S_7}) = 1}$. In this situation, the final choice of this voxel follows the highest probability of the corresponding voxel in the reference view, which is the $argmax({C_{1_7}})$. Therefore, the voxel ${C_{S_7}}$ is voted as the class 2.

\subsubsection{Weighted Averaging}

The weighted averaging is another fusion method adopted to realize the dynamical decision fusion. Different from the voting method, the weighted averaging is a linear operation process and mainly focuses on the influence of each result on the final one. It means that the segmentation result from multi-views makes their own contribution on the fusion segmentation result. To be more specific, the weighted averaging is defined as follows:

\begin{equation}
{O_{f_j}} = argmax({\sum\nolimits_{i = 1}^k {{\omega _i}{O_{i_j}}} }{\rm{) }}
\end{equation}

\begin{figure}[!ht]
\centering
\includegraphics[width=0.48\textwidth]{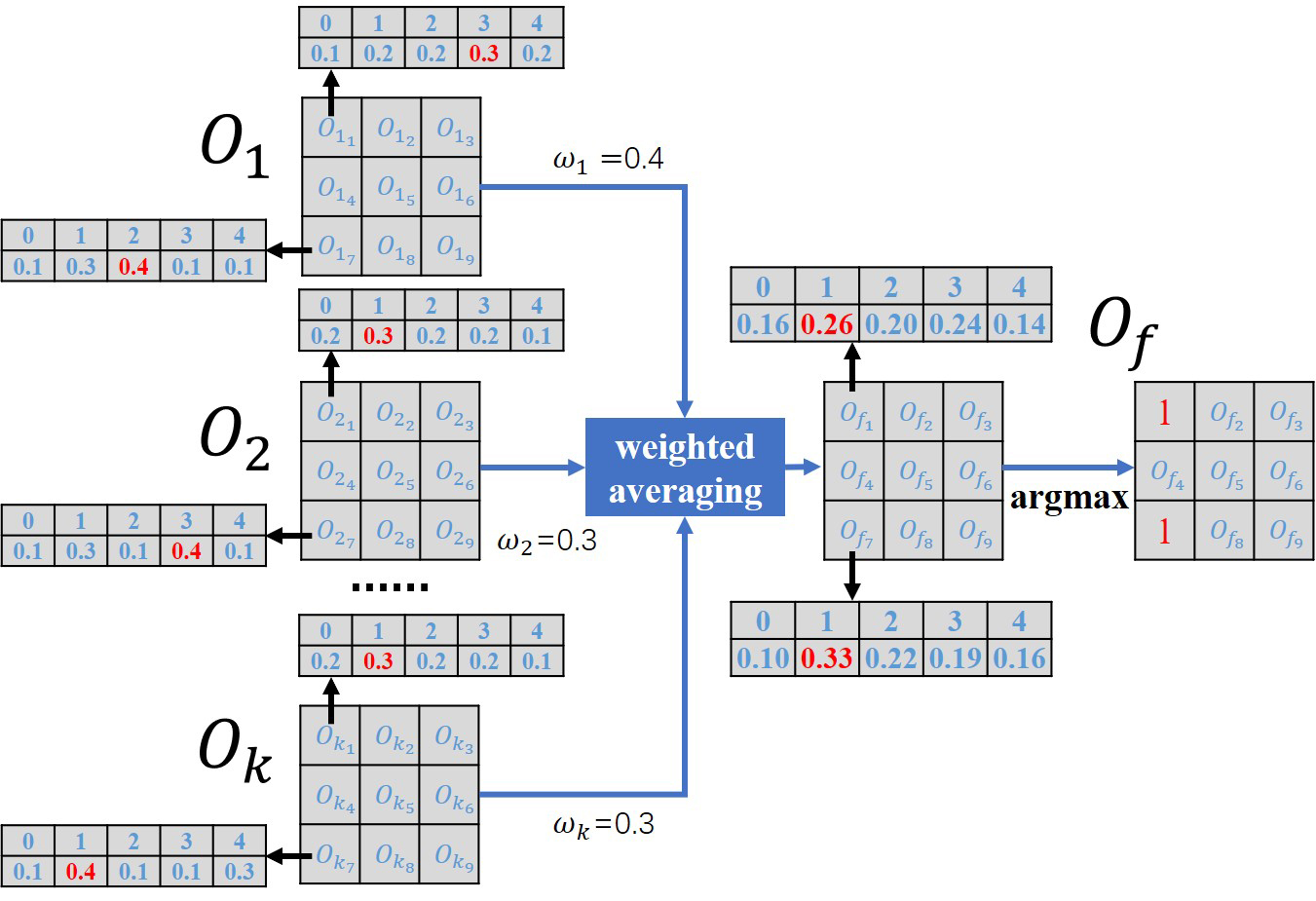}
\caption{An example of adopting the weighted averaging method to perform the data fusion.}
\label{fig:Weighted_averaging}
\end{figure}

Where $O_{f_j}$ represents the $j-th$ voxel in the final fusion segmentation result; $argmax()$ is used to achieve the class with the highest probability of one voxel to determine the final class; $\omega_i$ is the weighting factor corresponding to the $i-th$ view; $O_{i_j}$ indicates the $j-th$ voxel in the segmentation result from the $i-th$ view; the $k$ represents different views to observe the 3D dataset, where $ k \ge 1$ and $ k \le 3$. An example of the weighted averaging method to fuse segmentation results from multi-view is shown in Figure \ref{fig:Weighted_averaging}. The weighted averaging method firstly sets the weighting factors for different views. And for each voxel in different segmentation results from multi-views, the voxel with the same spatial position is fused by adopting the weighted averaging to obtain the class probability on the voxel in fusion result. For example, the probability of class 1 in $O_{f_1}$ is calculated by $0.4$ $(\omega_1)*0.2$ (the probability of class 1 in $O_{1_1}$) $+0.3$ $(\omega_2)*0.3$ (the probability of class 1 in $O_{2_1}$) $+0.3$ $(\omega_3)*0.3$ (the probability of class 1 in $O_{3_1}$), and the value is 0.26. After calculating each class probability of all voxels in the fusion result, the $argmax()$ is then adopted to determine the final class for the voxel. As shown in Figure \ref{fig:Weighted_averaging}, the $O_{f_1}$ is regarded as the class 1 which means that this voxel is a necrosis one. By adopting the weighted averaging method, the segmentation results from different views can be dynamically integrated as a fusion result and the better segmentation performance can be achieved by utilizing more comprehensive information from multi-views. 

\subsection{The Multi-View Fusion Loss} \label{section:Loss}

In this work, two types of segmentation results can be obtained from the proposed architecture, one is the output of multi-view network while another one is the fusion result. In order to facilitate the process of network training, the multi-view fusion loss is proposed to keep the consistency of appearance and space. The proposed multi-view fusion loss consists of the segmentation loss, the decision loss and the transition loss. The segmentation loss represents the loss between the segmentation result of the multi-view network from one single view and the Ground-Truth with the same view. Furthermore, the segmentation result not only represents the final output of learning network, but also indicates the result from the middle layer of network. It means the middle layer supervision method is also adopted in this work. The decision loss indicates the loss between the fusion segmentation result in one single view and the Ground-Truth with the corresponding view, while the transition loss is between the fusion segmentation result in one single view and the segmentation result of the multi-view network from the same view. 

Moreover, instead of directly adding these three losses, the weighted sum strategy is adopted to achieve a better performance. The multi-view fusion loss for the multi-view network is defined as follows:
 
\begin{equation}
{L_{multi-view{\ }fusion}} = {L_{segmentation}} + \alpha {L_{transition}} + \beta {L_{decision}}
\end{equation}

Where $\alpha$ and $\beta$ are the weighting factors of the transition loss and the decision loss, respectively. The multi-view fusion loss considers the interaction influence among the segmentation result from the multi-view network, the fusion segmentation result and the ground-truth. And it combines these losses as an integrate one to supervise the process of network training by adopting the back-propagation algorithm. 

\subsubsection{Segmentation Loss}

The segmentation loss is the most common loss to train the segmentation network. It is used to measure the difference between the segmentation result from the single view in the multi-view network and the Ground-Truth. Similar to the loss adopted in other deep learning methods, the cross-entropy function is employed as the basis to calculate the segmentation loss. Besides calculating the segmentation loss between the Ground-Truth and the final output of each network, the segmentation loss between the Ground-Truth and the output of the downsampling or upsampling stages in each network are also evaluated. Moreover, because of the extremely unbalanced class in the dataset, the weighted cross-entropy function(WCE) is eventually adopted in this paper, which is expressed as follows:

\begin{equation}
{L_{segmentation}} = - \frac{1}{T}\sum\limits_i^l {\sum\limits_{j = 1}^T {\sum\limits_{k = 1}^m {{\omega _k}{y_{{j_k}}}\log {O_{{i_{{j_k}}}}}}}} 
\end{equation}

Where $T$ is the size of the input image; $l$ is the output of different stage in each learning network, and in the downsampling stage of the FCN network the value of $l$ can be one of 1, 2, 3 and 4. While $l$ is set from 1 to 5 in the upsampling stage of the UNET network; $m$ represents the total number of class; $O_{i_{j_k}}$ represents the output probability of the k-th class of j-th voxel in the i-th stage; $y$ is the output of the One-Hot encoding when the Ground-Truth is adopted as the input; ${y_{j_k}}$ represents the Ground-Truth probability of the k-th class of the j-th voxel; $\omega$ is the weighting factor corresponding to the class, $\omega = [{\omega _1},{\omega _2}, \cdots ,{\omega _m}]$ is the set of weighting factor for each class in the prediction probability. It is mainly used to balance the influence of classes. In this way, the contribution of different classes with different proportions on the loss function becomes more balanced. The ${\omega _k}$ is computed as follows:

\begin{equation}
{\omega _k} = \frac{{T - \sum\nolimits_j^T {{O_{j_k}}}}}{{\sum\nolimits_j^T {{O_{j_k}}}}}
\end{equation}

\subsubsection{Transition Loss}

The proposed transition loss is mainly used to measure the difference between the segmentation result from the single view and the final fusion result from the corresponding view. The aim of transition loss is to provide more useful information from other views so as to guide the learning network in single view. Consequently, the fusion result can be used to dynamically optimize the network training process and translate the individually training process in different segmentation networks into a training process of mutual learning and assimilation. The calculation formula of the transition loss is expressed as follows:

\begin{equation}
{L_{transition}} = - \frac{1}{T}\sum\limits_{j = 1}^T {\sum\limits_{k = 1}^m {{\omega _k}{O_{f_{j_k}}}\log {O_{j_k}}}} 
\end{equation}

Where $T$ is the size of the input image; $m$ is the total number of class; $O_{f_{j_k}}$ represents the fusion probability of the j-th voxel's k-th class, and $O_{j_k}$ represents the network output probability of the j-th voxel's k-th class from a specific view before applying the decision fusion method.

\subsubsection{Decision Loss}

The proposed decision loss is mainly used to measure the difference between the fusion segmentation result from the single view and the Ground-Truth with the corresponding view. Its purpose is to make the fusion segmentation result to get closer to the ground-truth and to guide the multi-view learning network to better learn useful features from the input images from multi-views. The calculation formula is defined as follows:

\begin{equation}
{L_{decision}} = - \frac{1}{T}\sum\limits_{j = 1}^T {\sum\limits_{k = 1}^m {{\omega _k}{y_{j_k}}\log {O_{f_{j_k}}}} } 
\end{equation}

Where $T$ is the size of the input image; $m$ is the total number of class; ${y_{j_k}}$ represents the Ground-Truth's probability of the j-th voxel's k-th class, and $O_{f_{j_k}}$ represents the fusion's probability of the j-th voxel's k-th class. 

\section{Experiments, Results, and Discussion} \label{section:Experiments, Results, and Discussion}

\subsection{Experimental Details}

\subsubsection{Dataset}

Our experiments are mainly evaluated on the Brain Tumor Image Segmentation Challenge 2015 (BRATS 2015). Ample multi-institutional routine clinically-acquired pre-operative multi-modal MRI scans of glioblastoma (GBM/HGG) and lower grade glioma (LGG) are provided as the training and testing data for BRATS 2015 challenge. The training dataset contains 220 patients in HGG and 54 patients in LGG. Each patient has four different modalities of MRI, which are T2-weighted fluid attenuated inversion recovery (Flair), T1-weighted (T1), T1-weighted contrast-enhanced (T1c), and T2-Weighted (T2). Multi-modal MRI images are mainly used to analyze the pathological feature of brain tumors. The online testing dataset consists of MRI images of 110 patients both in HGG and LGG. And the Ground Truth of the online test dataset is not publicly available. Each modal for a patient is a 3D data with the size of $155*240*240$. The task is to segment these multi-modal brain tumor image into five classes: necrosis, edema, non-enhancing tumor, enhanced tumor, and everything else.

\subsubsection{Local Validation}
In the experiment, the 220 patients' data from HGG in the training dataset is further divided into two parts: the first 190 patients' data is also adopted as the training dataset; and the last 30 patients' data is employed as the validation dataset. When evaluating the effectiveness of the proposed method on the validation dataset, the FCN network is adopted as the main learning network in the multi-view deep neural network architecture. It should be noted that the experimental results on the validation dataset usually perform better than the testing dataset. The reason is that there is the over-fitting problem for the validation dataset. In general, the validation dataset is mainly used to verify the effectiveness of proposed framework. While the testing dataset is used to evaluate the eventual performance of the proposed multi-view framework.  

\subsubsection{Online Testing}
In this work, the online testing dataset in BRATS 2015 is adopted to evaluate the proposed method. The testing dataset includes the HGG, the LGG and some synthetic data of 110 patients. Compared with the testing dataset in BRATS 2013, the BRATS 2015 is a more challenging one. And the performance is usually degraded when segmentation methods evaluate both on the BRATS 2013 and the BRATS 2015. The proposed framework is firstly trained with the training dataset and then is used to segment the brain tumor on the testing dataset. These segmentation results are uploaded to the official website to obtain the final segmentation performance.  

\subsubsection{Evalution}

The evaluation metric is calculated on 3 different tumor sub-compartments: the complete tumor region, the core tumor region, and the enhancing tumor region. The complete tumor region is composed of all four tumor classes. The core tumor region contains all tumor classes except the edema class, while the enhancing tumor region only contains enhanced tumor class. The dice coefficient is adopted to evaluate the segmentation performance on these three tumor regions and is defined as follows:

\begin{equation}
Dice(GT,AT) = \frac{{2|GT \cap AT|}}{{|GT| + |AT|}}
\end{equation}

Where the $GT$ represents the ground truth and the $AT$ represents the predictions from the proposed method.

\subsubsection{Training Details}

The multi-view network is used to segment the brain tumor from the axial, coronal and sagittal views. In the process of network training, the initial learning rate of the network is set to 0.0001, and it is reduced by half every 2 epochs during network training. The dropout strategy is also adopted and the parameter is set to 0.2 during network training. The size of each patient's data is expanded from $155*240*240$ to $160*240*240$ by filling the 0. When training the individual network just in a single view, the image's batch size of the input is set to 16. While training the multi-view network, the input image's batch size is set to 16, 16 and 10 corresponding to the coronal, sagittal, and axial view respectively. And in the last image batch for each patient's data, this value is changed to 20 in the axial view. The reason behind these settings is the number of input images for learning network are different from multi-views. These different settings for input image batches ensure that different learning networks corresponding to multi-views can simultaneously obtain the network output for future fusion in the process of dynamic decision fusion. Moreover, all experiments are run on the Nvidia GTX 2080ti. It costs about 8 hours to train the learning network for a single view with 35 epochs. \par

\subsection{The Segmentation Performance on Single View}

As a comparison, the segmentation performances on the coronal view, sagittal view and axial view are firstly evaluated. The FCN network is adopted as the backbone network of the multi-view network architecture to realize the brain tumor segmentation. After training these three segmentation networks, the final segmentation performance is evaluated on the validation dataset and the experimental results are shown in Table \ref{tab:different_views}. It can be found that the segmentation from the axial view achieves the best performance with a dice metric of 0.879 in complete tumor region, 0.827 in the core tumor region, and 0.794 in the enhancing tumor regions. While the segmentation from the coronal view is the second best by following with the segmentation result from the sagittal view. But the biggest difference on segmentation results from these views is mainly reflected in the complete tumor region and the core tumor region. Moreover, it also reveals why most segmentation methods only evaluate the image from the axial view without considering the images from other two views. One possible reason is that the image from the axial view is relatively symmetrical and demonstrates the most useful information of the brain tumor. However, besides the axial view, it can be said that the other two views can also facilitate the process of segmenting the brain tumor. 

\begin{table}[!htbp]
  \centering
  \caption{The segmentation performance on single view method by adopting the FCN network on the validation dataset.}
    \begin{tabular}{cccccccc}
    \toprule
    \multirow{2}[4]{*}{network} & \multirow{2}[4]{*}{dataset} & \multicolumn{3}{c|}{view} & \multicolumn{3}{c}{dice} \\
\cmidrule{3-8}          &       & axial & coronal & \multicolumn{1}{c|}{sagittal} & complete &  core  & enhancing  \\
    \midrule
    \multirow{3}[2]{*}{FCN} 
    & \multicolumn{1}{c}{\multirow{3}[2]{*}{validation}} & $\surd$     & $\times$      &  $\times$     & \textbf{0.879}  & \textbf{0.827}  & \textbf{0.794}  \\
          &       & $\times$      & $\surd$     &  $\times$     & 0.857  & 0.814  & 0.788  \\
          &       & $\times$      & $\times$      & $\surd$     & 0.853  & 0.795  & 0.789  \\
    \bottomrule
    \end{tabular}
  \label{tab:different_views}
\end{table}

\subsection{The Segmentation Performance on Multi-view Fusion}

In this section, the segmentation performance by fusing results from multi-views is evaluated to improve the brain tumor segmentation performance. Two types of dynamic fusion methods, the voting method and the weighted averaging method, are adopted to dynamically fuse the segmentation results from multi-views. By evaluating these two fusion methods on the validation dataset, the experimental results can be found in Table \ref{tab:different_fusion}. \par 

\subsubsection{Voting Method}
The voting method is used to fuse the segmentation results from at least three views and also needs a reference view. In the experiment, the coronal view, sagittal view and axial view are regarded as a reference view respectively for evaluating the voting method. As seen in Table \ref{tab:different_fusion}, the voting method of referring the axial view achieves the best performance when adopting the voting dynamic fusion method. Furthermore, all three voting methods achieve a better segmentation performance compared to the segmentation result only employing the single view. To be more specific, the voting method referring the axial view raises all three dice metrics and the enhancing tumor by 1\% and 2\% respectively compared to the best segmentation performance on adopting the single view (from the axial view). The voting method referring the coronal view and the sagittal view also achieves at least 1\% improvement on the segmentation performance. It can be proven that the multi-views fusion method holds great potential to improve the brain tumor segmentation performance compared to the segmentation method only considering the image from the single view. \par

\begin{table*}[!hbp]
  \centering
  \caption{The segmentation performance on multi-view dynamic fusion method by adopting the voting method and the weighted averaging method on the validation dataset.}
  \resizebox{0.99 \textwidth}{!}{
    \begin{tabular}{cccccccccc}
    \toprule
    \multirow{2}[4]{*}{network} & \multirow{2}[4]{*}{dataset} & \multicolumn{3}{c}{view} & \multirow{2}[4]{*}{fusion method} & \multirow{2}[4]{*}{reference} & \multicolumn{3}{c}{dice} \\
\cmidrule{3-5}\cmidrule{8-10}          &       & axial & coronal & sagittal &       &       & complete &  core  & enhancing  \\
    \midrule
    \multirow{8}[2]{*}{FCN} 
    & \multicolumn{1}{c}{\multirow{8}[2]{*}{validation}} & $\surd$     & $\times$      & $\times$      & -     & -     & 0.879  & 0.827  & 0.794  \\
          &       & $\surd$     & $\surd$     & $\surd$     & voting & axial & 0.893  & 0.841  & 0.812  \\
          &       & $\surd$     & $\surd$     & $\surd$     & voting & coronal & 0.892  & 0.838  & 0.811  \\
          &       & $\surd$     & $\surd$     & $\surd$     &  voting & sagittal & 0.891  & 0.833  & 0.808  \\
          &       & $\surd$     & $\surd$     & $\times$      & weighted averaging & -     & 0.896  & 0.844  & 0.811  \\
          &       & $\surd$     & $\times$      & $\surd$     & weighted averaging & -     & 0.899  & 0.842  & 0.806  \\
          &       & $\times$      & $\surd$     & $\surd$     & weighted averaging & -     & 0.891  & 0.832  & 0.813  \\
          &       & $\surd$     & $\surd$     & $\surd$     & weighted averaging & -     & \textbf{0.901}  & \textbf{0.847}  & \textbf{0.825}  \\
    \bottomrule
    \end{tabular}
    }
  \label{tab:different_fusion}
\end{table*}

\subsubsection{Weighted averaging}
Besides the voting method, the weighted averaging is also adopted to facilitate the process of dynamic decision fusion in the proposed framework. Different from the voting method, the weighted averaging can be used to fuse the segmentation results from two or more views. For the segmentation results from the axial view, coronal view and sagittal view, there are four different fusion combinations, where three combinations are based on the fusion of two views and one combination is the fusion of three views. In the experiment, if the weighted averaging is used to fuse the segmentation results from two views, the weighting factors are set to 0.5 and 0.5. And the weighting factors are set to 0.4, 0.3 and 0.3 for the axial view, coronal view and saggittal view respectively when fusing segmentation results from three views. As seen in Table \ref{tab:different_fusion}, it can be found that all weighted averaging methods (fusing two or three views) obtain a better performance compared to the voting methods and the segmentation result only adopting the single view. To be more specific, the weighted averaging method of fusing coronal view and sagittal view achieves a minimum improvement on the segmentation performance compared to other weighted averaging methods of fusing the axial view and other views. According to this, it can be proven that the axial view contains the most useful information for segmenting the brain tumor. Furthermore, compared to the weighted averaging method of only fusing two views, the method of fusing three views achieves the best segmentation performance with the dice metric of 0.901 in the complete region, 0.847 in the core region, and 0.825 in the enhancing region. It raises all three dice metrics and enhancing tumor by 2\% and 3\% respectively compared to the best segmentation result obtained from the single view. Moreover, it can be proven again that the proposed multi-views method has the ability to facilitate the process of segmenting the brain tumor and improve the segmentation performance by employing more useful information from multi-views. 

\begin{figure*}[!htbp]
\centering
\includegraphics[width=\textwidth]{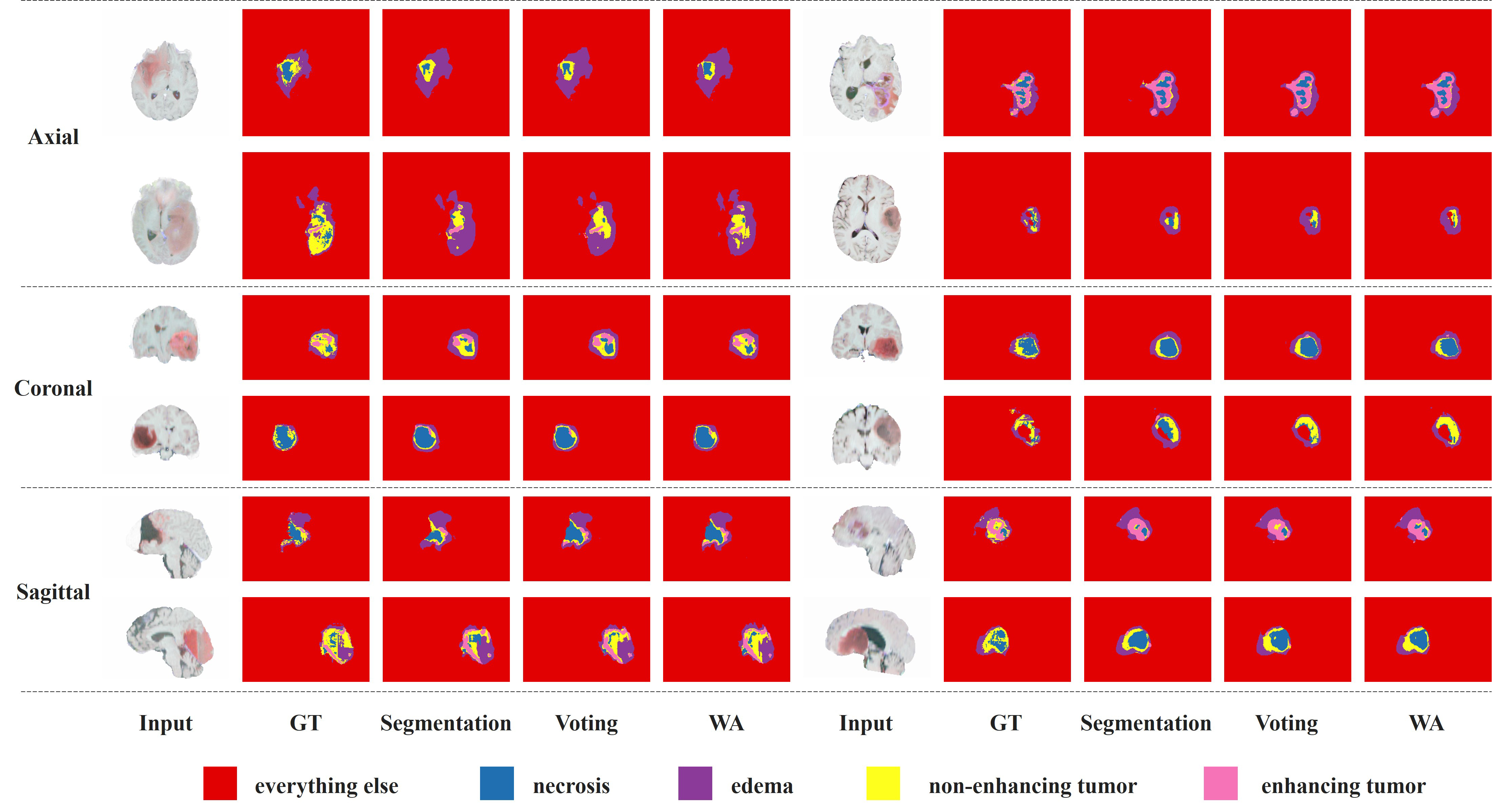}
\caption{Visually comparing the segmentation results of the two decision fusion methods based on the FCN segmentation network in the validation dataset. From left to right are: Input: Input images connected by four modalities data; GT: the Ground-Truth; Segmentation: Original outputs of the network; Voting: The multi-view fusion result based on the voting method of referring the axial view; WA: The multi-view fusion results based on the weighted averaging method of fusing three views.}
\label{fig:Segmentation}
\end{figure*}

\begin{table*}[!hb]
  \centering
  \caption{The segmentation performance on the multi-view dynamic fusion method with/without the multi-view fusion loss on the validation dataset.}
   \resizebox{0.99 \textwidth}{!}{ 
    \begin{tabular}{ccccccccccc}
    \toprule
    \multirow{2}[4]{*}{network} & \multirow{2}[4]{*}{dataset} & \multicolumn{3}{c}{view} & \multicolumn{1}{c}{\multirow{2}[4]{*}{fusion loss}} & \multirow{2}[4]{*}{fusion method} & \multirow{2}[4]{*}{reference} & \multicolumn{3}{c}{dice} \\
\cmidrule{3-5}\cmidrule{9-11}          &       & \multicolumn{1}{c}{axial} & \multicolumn{1}{c}{coronal} & \multicolumn{1}{c}{sagittal} & \multicolumn{1}{c}{} &       &       & complete &  core  & enhancing  \\
    \midrule
    \multirow{8}[2]{*}{FCN} & \multicolumn{1}{c}{\multirow{8}[2]{*}{validation}} & $\surd$  & $\surd$  & $\surd$  & $\times$ & voting & axial & 0.893  & 0.841  & 0.812  \\
          &       & $\surd$  & $\surd$  & $\surd$  & $\surd$  & voting & axial & 0.899  & 0.843  & 0.814  \\
          &       & $\surd$  & $\surd$  & $\surd$  & $\times$ & voting & coronal & 0.892  & 0.838  & 0.811  \\
          &       & $\surd$  & $\surd$  & $\surd$  & $\surd$  & voting & coronal & 0.897  & 0.844  & 0.813  \\
          &       & $\surd$  & $\surd$  & $\surd$  & $\times$ & voting & sagittal & 0.891  & 0.833  & 0.808  \\
          &       & $\surd$  & $\surd$  & $\surd$  & $\surd$  & voting & sagittal & 0.898  & 0.840  & 0.810  \\
          &       & $\surd$  & $\surd$  & $\surd$  & $\times$ & weighted averaging & -     & \textbf{0.901}  & 0.847  & 0.825  \\
          &       & $\surd$  & $\surd$  & $\surd$  & $\surd$  & weighted averaging & -     & \textbf{0.901} & \textbf{0.850}  & \textbf{0.829}  \\
    \bottomrule
    \end{tabular}
    }
  \label{tab:loss}
\end{table*}

Figure \ref{fig:Segmentation} shows the comparison of the fusion of multi-views with different dynamic decision fusion methods on the validation dataset in a visual way. Each two rows represents the brain tumor segmentation result from the axial view, coronal view and sagittal view by adopting the single view method, the voting multi-views fusion method and the weighted averaging multi-views fusion method, respectively. As seen from Figure \ref{fig:Segmentation}, it can be found that the segmentation performance from the multi-views methods is better than the single view method and the segmentation result based on the weighted averaging method is most close to the ground truth. It is also verified that the information from multi views has the ability to improve the final segmentation result.

\subsection{Evaluating the Effectiveness of the Multi-View Fusion Loss}

The proposed multi-view fusion loss is mainly used to keep the consistency of appearance and space when adopting the multi-view method to segment the brain tumor. In this experiment, the effectiveness of the proposed multi-view fusion loss is evaluated to facilitate the training process of multi-view network. The multi-view fusion loss consists of the segmentation loss, decision loss and transition loss. The weighting factor of transition loss $\alpha$ is set to 0.5 while the weighting factor of decision loss $\beta$ is set to 1. The FCN network is adopted as the learning network for the multi-view network architecture, while the dynamic decision fusion method employs both the voting method and the weighted averaging method of fusing three views. The Table \ref{tab:loss} shows the evaluation results of the proposed multi-view fusion loss on the validation dataset. It can be seen that the segmentation performance is improved and the dice metric is raised in all three tumor regions by adopting the proposed multi-view fusion loss for the voting method and the weighted averaging method. More specifically, when adopting the proposed multi-view fusion loss on the voting methods, the segmentation performance on the complete tumor region obtains a comparative improvement with approximately 0.6\% increase compared to the voting method without employing the multi-view fusion loss. The weighted averaging method with the multi-view fusion loss achieves a slight improvement on the dice metric of the core tumor region and the enhancing tumor region, which is raised by approximately 0.3\% to 0.4\%. Overall, the weighted averaging method of fusing three views, which also adopts the multi-view fusion loss achieves the best segmentation performance with the dice metric of 0.901 in the complete tumor region, 0.850 in the core tumor region, and 0.829 in the enhancing tumor region. By conducting experiments about the multi-view fusion loss, it can be proven that the proposed loss has the ability to facilitate the learning network training process and it is an effectiveness improvement to obtain a better brain tumor segmentation performance. It can be explained that there is a strong relationship between the fusion segmentation result and the segmentation result obtained from the multi-view network. This is why the transition loss is proposed to indicate the loss between these two segmentation results. Instead of affecting the process of data fusion from multi-view, this transition loss is directly used to facilitate the multi-view network training by adopting the back-propagation method, which can make each output of the network closer to the fusion result. Meanwhile, the decision loss is used to make fusion segmentation result closer to the Ground Truth, which also can directly facilitate the learning network training process. 
\par

\begin{table*}[!hbp]
  \centering
  \caption{The segmentation performance of adopting the FCN network and UNET network as the backbone network of multi-view dynamic fusion method on the testing dataset.}
    \resizebox{0.99 \textwidth}{!}{ 
    \begin{tabular}{ccccccccccc}
    \toprule
    \multirow{2}[4]{*}{network} & \multirow{2}[4]{*}{dataset} & \multicolumn{3}{c}{view} & \multicolumn{1}{c}{\multirow{2}[4]{*}{fusion loss}} & \multirow{2}[4]{*}{fusion method} & \multirow{2}[4]{*}{reference} & \multicolumn{3}{c}{dice} \\
\cmidrule{3-5}\cmidrule{9-11}          &       & \multicolumn{1}{c}{axial} & \multicolumn{1}{c}{coronal} & \multicolumn{1}{c}{sagittal} & \multicolumn{1}{c}{} &       &       & complete & core  & enhancing \\
    \midrule
    \multirow{3}[2]{*}{FCN} & \multicolumn{1}{c}{\multirow{3}[2]{*}{testing}} & $\surd$ & $\times$ & $\times$ & $\times$ & -     & -     & 0.81  & 0.65  & 0.52  \\
          &       & $\surd$ & $\surd$ & $\surd$ & $\surd$ &  voting & axial & 0.82  & 0.66  & 0.55  \\
          &       & $\surd$ & $\surd$ & $\surd$ & $\surd$ & weighted averaging & -     & 0.83  & 0.66  & 0.57  \\
    \midrule
    \multirow{3}[2]{*}{UNET} & \multicolumn{1}{c}{\multirow{3}[2]{*}{testing}} & $\surd$ & $\times$ & $\times$ & $\times$ & -     & -     & 0.82  & 0.67  & 0.57  \\
          &       & $\surd$ & $\surd$ & $\surd$ & $\surd$ & voting & axial & \textbf{0.84}  & 0.67  & 0.59  \\
          &       & $\surd$ & $\surd$ & $\surd$ & $\surd$ & weighted averaging & -     & \textbf{0.84}  & \textbf{0.68}  & \textbf{0.61}  \\
    \bottomrule
    \end{tabular}
    }
  \label{tab:different_backbone}
\end{table*}

\begin{table*}[!hbp]
  \centering
  \caption{The segmentation performance of multi-view dynamic fusion method based on the UNET network on the testing dataset.}
    \resizebox{0.99 \textwidth}{!}{ 
    \begin{tabular}{ccccccccccc}
    \toprule
    \multirow{2}[4]{*}{network} & \multirow{2}[4]{*}{dataset} & \multicolumn{3}{c}{view} & \multirow{2}[4]{*}{fusion loss} & \multirow{2}[4]{*}{fusion method} & \multirow{2}[4]{*}{reference} & \multicolumn{3}{c}{dice} \\
\cmidrule{3-5}\cmidrule{9-11}          &       & axial & coronal & sagittal &       &       &       & complete & core  & enhancing \\
    \midrule
    \multirow{14}[2]{*}{UNET} & \multirow{14}[2]{*}{testing} & $\surd$ & $\times$ & $\times$ & $\times$ & -     & -     & 0.82  & 0.67  & 0.57  \\
          &       & $\times$ & $\surd$ & $\times$ & $\times$ & -     & -     & 0.82  & 0.64  & 0.56  \\
          &       & $\times$ & $\times$ & $\surd$ & $\times$ & -     & -     & 0.80  & 0.63  & 0.57  \\
          &       & $\surd$ & $\surd$ & $\surd$ & $\times$ & voting & axial & \textbf{0.84}  & 0.67  & 0.58  \\
          &       & $\surd$ & $\surd$ & $\surd$ & $\surd$ & voting & axial & \textbf{0.84}  & 0.67  & 0.59  \\
          &       & $\surd$ & $\surd$ & $\surd$ & $\times$ & voting & coronal & \textbf{0.84}  & 0.67  & 0.58  \\
          &       & $\surd$ & $\surd$ & $\surd$ & $\surd$ & voting & coronal & \textbf{0.84}  & 0.67  & 0.58  \\
          &       & $\surd$ & $\surd$ & $\surd$ & $\times$ & voting & sagittal & \textbf{0.84}  & 0.66  & 0.58  \\
          &       & $\surd$ & $\surd$ & $\surd$ & $\surd$ & voting & sagittal &\textbf{0.84}  & 0.66  & 0.58  \\
          &       & $\surd$ & $\surd$ & $\times$ & $\times$ & weighted averaging & -     & \textbf{0.84}  & 0.66  & 0.59  \\
          &       & $\surd$ & $\times$ & $\surd$ & $\times$ & weighted averaging & -     & 0.83  & 0.66  & 0.58  \\
          &       & $\times$ & $\surd$ & $\surd$ & $\times$ & weighted averaging & -     & 0.83  & 0.65  & 0.59  \\
          &       & $\surd$ & $\surd$ & $\surd$ & $\times$ & weighted averaging & -     & \textbf{0.84}  & 0.67  & 0.60  \\
          &       & $\surd$ & $\surd$ & $\surd$ & $\surd$ & weighted averaging & -     & \textbf{0.84}  & \textbf{0.68}  & \textbf{0.61}  \\
    \bottomrule
    \end{tabular}
    }
  \label{tab:U_testing}
\end{table*}

\subsection{Evaluating the Effectiveness of Different Backbone Network}

The aforementioned experiments prove the effectiveness of the proposed multi-view dynamic fusion framework by conducting the experiments on the validation dataset. The FCN network is adopted as the learning network to extract the features and segment the brain tumor from different views. The FCN network structure is relatively simple and it is a good choice to firstly verify the proposed multi-view method. However, the UNET network is widely adopted in the area of medical image segmentation and it can achieve a better performance than the FCN network \cite{ref21,ref38,ref40}. In order to further improve the brain tumor segmentation performance, the UNET network and the FCN network are respectively adopted as the backbone network to realize the multi-view network architecture. And instead of the validation dataset, the testing dataset is employed to evaluate these two backbone networks so as to achieve the state-of-the-art segmentation performance. \par

In this experiment, for the FCN network, we only chose the best performance in the single view, the multi-views with voting fusion method and the multi-views with weighted averaging fusion method as the comparison. As shown in Table \ref{tab:different_backbone}, it can be found that the UNET network achieves a better segmentation performance both on the single view method and the multi-view fusion method compared to the FCN network. More specifically, in the segmentation result of the single view, the UNET raises the dice metric by approximately 1\%, 2\% and 5\% in the complete, core, and enhancing tumor regions, respectively. While adopting the multi-view fusion method with the voting and weighted averaging, the UNET backbone network also achieves approximately 1\%, 1\% and 4\% promotion in the complete, core and enhancing tumor regions. The reason behind this is that the UNET network is with a better feature extraction capability for medical images. Moreover, it can be proven that the UNET is a better backbone network to realize the multi-view network architecture and has the ability to achieve a better brain tumor segmentation performance. In the following experiment, the UNET network is adopted as the backbone network to implement the proposed multi-view fusion method and the testing dataset is used to conduct the experiment to obtain the convincing segmentation results. 

Furthermore, the experiments on the single view, the voting fusion method, the weighted averaging fusion method and the multi-view fusion loss are also conducted on the UNET network by employing the testing dataset. The Table \ref{tab:U_testing} presents all these experimental results. It can be found the segmentation result from the axial view obtains the best performance in segmentation methods of only employing the single view. Moreover, all fusion methods of fusing segmentation results from multi-views always achieve a better performance compared to the single view method. In the voting fusion method, the method which refers the axial view is the best one. But the averaging weighted method always obtains a better segmentation performance than the voting fusion method. While the weighted averaging method fusing three views is better than the ones just fusing two views. Furthermore, the multi-view fusion loss has the ability to further improve the segmentation performance. Overall, it can be noted that the best segmentation performance is raised to 0.84, 0.68 and 0.61 for the complete tumor region, the core tumor region and the enhancing tumor region, respectively by evaluating the proposed method on the testing dataset. It can be proven that when evaluating the proposed method on the testing dataset to obtain the final segmentation performance, the UNET network is a better choice than the FCN network to realize the proposed multi-view method. In the following experiments, the UNET network is adopted as the backbone network to evaluate other improvement on the testing dataset.

\subsection{Evaluating Different Parameter Settings for the Proposed Multi-View Fusion Loss}

In the previous experiment, the proposed multi-view fusion loss has been evaluated. In the multi-view fusion loss, the segmentation loss is the most common one, which is widely used for training the deep learning network. Therefore, we just discuss the influence of different parameter settings for the proposed transition loss and decision loss, which includes the weighting factors and the time of employing the loss. In this experiment, the backbone network is the UNET, and the fusion method employs the weighted averaging of fusing three views. All experiments are evaluated on the testing dataset. \par

\begin{table}[!h]
  \centering
  \caption{The comparison of different weighting factors for the transition loss and the decision loss when adopting the weighted averaging fusion method on the testing dataset. $\alpha$ is the weighting factor of the transition loss and $\beta$ is the weighting factor of the decision loss, and the fusion loss is added to the total network loss after the beginning of the third epoch of the network training.}
    \begin{tabular}{ccccc}
    \toprule
    \multirow{2}[4]{*}{$\alpha$} & \multirow{2}[4]{*}{$\beta$} & \multicolumn{3}{c}{dice} \\
\cmidrule{3-5}          &       & complete & core  & enhancing \\
    \midrule
    0     & 0     & 0.84  & 0.67  & 0.60  \\
    0.5   & 0     & 0.84  & \textbf{0.68} & 0.60  \\
    0     & 1.0   & 0.84  & 0.67  & \textbf{0.61} \\
    0.5   & 0.5   & 0.84  & \textbf{0.68} & 0.60  \\
    1.0   & 1.0   & 0.84  & 0.67  & 0.60  \\
    1.5   & 1.0   & 0.84  & 0.67  & 0.60  \\
    0.5   & 1.0   & 0.84  & \textbf{0.68} & \textbf{0.61} \\
    0.5   & 1.5   & 0.84  & \textbf{0.68} & 0.60  \\
    \bottomrule
    \end{tabular}
  \label{tab:loss_factors}
\end{table}

The effectiveness of different weighting factors for the transition loss and the decision loss are firstly evaluated. In order to avoid the mutual inference with each other, the transition loss and the decision loss are constantly employed at the beginning of the third epoch in the network training so as to discuss the impact of different weighting factors on the transition loss and decision loss. The experimental results are shown in Table \ref{tab:loss_factors}. $\alpha$ is the weighting factor of the transition loss and $\beta$ is the weighting factor of the decision loss. Based on the control variable analysis, it can be found that the segmentation performance with different parameters of $\alpha$ and $\beta$ achieves a slight difference on the core region and the enhance region. More specifically, compared to the situation of not employing the transition loss and decision loss, if only employing the transition loss (the $\beta$ is set to 0), the dice metric of the core tumor region is increased by approximately 1\%. When only the decision loss is employed (the $\alpha$ is set to 0), the dice metric of the enhancing tumor region obtains approximately 1\% increase. It can be proven that the transition loss and the decision loss both are effective to improve the brain tumor segmentation. The best segmentation performance is achieved when $\alpha$ = 0.5 and $\beta$ = 1, where the dice metric in the complete, core, and enhancing tumor regions are 0.84, 0.68 and 0.61, respectively. \par

It can be explained that the transition loss is used to make the multi-view network segmentation result closer to the fusion segmentation results, while the decision loss is used to make fusion segmentation result closer to the Ground Truth. The ultimate goal of the proposed framework is to achieve the best segmentation performance as well as the Ground-Truth. The decision loss should make more important contribution to the final result. Consequently, the weighting factor of the transition loss should be set with a smaller value and the weighting factor of the decision loss can be set with a lager one. To be more specific, the best setting for the weighting factor of the decision loss is the same as the weighting factor of segmentation loss (the default value is 1), which means that the fusion segmentation result owns the same important influence to facilitate the network training compared to the segmentation result obtained from the single view. \par

Besides the weighting factor of the transition loss and the decision loss, when to employ the proposed loss also has the influence on the final segmentation performance. In this experiment, we conduct experiments to find the best time for employing the transition loss and decision loss during the multi-view network training. The epoch of network training is regarded as the time to employ the transition loss and decision loss. The backbone network is the UNET network, and the weighting factors $\alpha$ and $\beta$ are fixed into the 0.5 and 1.0. For different time settings of employing the proposed loss, the network is trained with 35 epochs for each time. All experiments are evaluated on the testing dataset and the experimental results are shown in Table \ref{tab:loss_timing}. \par

\begin{table}[!htbp]
  \centering
  \caption{The Comparison of different time to employ the multi-view fusion loss by adopting the UNET network on the testing dataset. The beginning epoch represents the fusion loss is added at the beginning of the epoch of network training, and the weighting factors of the transition loss and the decision loss are set to 0.5 and 1, respectively.}
    \begin{tabular}{cccc}
    \toprule
    \multirow{2}[4]{*}{epoch to start} & \multicolumn{3}{c}{dice} \\
\cmidrule{2-4}          & complete & core  & enhancing \\
    \midrule
    1     & \textbf{0.85} & 0.67  & 0.60  \\
    3     & 0.84  & \textbf{0.68} & \textbf{0.61} \\
    5     & 0.84  & 0.67  & \textbf{0.61} \\
    7     & 0.84  & 0.67  & \textbf{0.61} \\
    \bottomrule
    \end{tabular}
  \label{tab:loss_timing}
\end{table}

\begin{table*}[hb]
  \centering
  \caption{The segmentation performance of adopting other improvements by adopting the UNET network on the testing dataset in BRATS 2015 challenge.}
  \resizebox{0.99 \textwidth}{!}{ 
    \begin{tabular}{ccccccccccc}
    \toprule
    \multicolumn{3}{c}{view} & \multirow{2}[4]{*}{fusion loss} & \multirow{2}[4]{*}{fusion method} & \multirow{2}[4]{*}{modal} & \multirow{2}[4]{*}{smaller image patches} & \multirow{2}[4]{*}{post-processing} & \multicolumn{3}{c}{dice} \\
\cmidrule{1-3}\cmidrule{9-11}    axial & coronal & sagittal &       &       &       &       &       & complete & core  & enhancing \\
    \midrule
    $\surd$ & $\times$ & $\times$ & $\times$ & $\times$ & Flair, T1c, T2 & $\times$ & $\times$ & 0.83  & 0.66  & 0.56  \\
    $\surd$ & $\times$ & $\times$ & $\times$ & $\times$ & Flair, T1, T1c, T1 & $\times$ & $\times$ & 0.82  & 0.67  & 0.57  \\
    $\surd$ & $\times$ & $\times$ & $\times$ & $\times$ & Flair, T1, T1c, T2 & $\surd$ & $\times$ & 0.83  & 0.69  & 0.58  \\
    $\surd$ & $\surd$ & $\surd$ & $\surd$ & weighted averaging & Flair, T1, T1c, T2 & $\times$ & $\times$ & 0.84  & 0.68  & \textbf{0.61}  \\
    $\surd$ & $\surd$ & $\surd$ & $\surd$ & weighted averaging & Flair, T1, T1c, T2 & $\surd$ & $\times$ & \textbf{0.85}  & 0.69  & \textbf{0.61}  \\
    $\surd$ & $\surd$ & $\surd$ & $\surd$ & weighted averaging & Flair, T1, T1c, T2 & $\surd$ & $\surd$ & \textbf{0.85}  & \textbf{0.71}  & \textbf{0.61}  \\
    \bottomrule
    \end{tabular}
    }
  \label{tab:others_improvements}
\end{table*}

On account of the situation of employing the transition loss and decision loss at the beginning of the network training, at this moment, each learning network corresponding to the axial view, coronal view and sagittal view doesn't reach a stable state in the early epoch of network training. Although the dice metric of the complete tumor region is 0.85, the performance on the core and enhancing regions is not good enough. When adopting the proposed loss at the fifth or seventh epoch, the dice metric of enhancing region achieves a 1\% promotion to 0.61. But the dice metric of complete region degrades to 0.84 compared to employing the proposed loss at the beginning of the network training. When all improvements are considered together, the best performance is achieved by employing the transition loss and decision loss at the third epoch of the network training, where the dice metric is 0.84, 0.68 and 0.61 on the complete, core and enhancing tumor region. More importantly, the transition loss and the decision loss are worked on the fusion segmentation result and are mainly used to keep the consistency of the appearance and space during network training. When adopting the transition loss and the decision loss at the beginning of network training, each segmentation network just considers the segmentation problem from the single view and lacks the useful information from multi-views. Therefore, the proposed transition loss and decision loss can't take advantage of the fusion information to train the multi-view network. However, if the proposed loss is adopted at the later stage of network training, the fusion result will produce an influence large enough on the multi-view network. And the multi-view network has learned a lot of information from the fusion result, no matter whether the information is right or wrong. Therefore, the proposed loss is adopted too late to facilitate the multi-view network training. Overall, the best choice is to adopt the transition loss and the decision loss at the third epoch of network training. Moreover, according to this experiment, it is also proven that the proposed multi-view fusion loss is an effective improvement to achieve a better brain tumor segmentation from another side.

\subsection{Evaluating the Effectiveness of Other Improvements}

In order to further improve the proposed multi-view dynamic fusion framework and achieve a better segmentation performance, some simple yet effective improvements are proposed. \cite{ref17} found that a brain tumor segmentation model only adopting the Flair, T1c and T2 model as the input can achieve a better performance than the model which employs the whole four modalities Flair, T1, T1c and T2 as the input. Based on the founding in \cite{ref17}, an experiment has been evaluated to show the segmentation performance only employing three modalities as the network input in the axial view. As shown in the first and second rows of the Table \ref{tab:others_improvements}, the dice metrics of the complete, core and enhancing tumor regions obtained by the three modalities are 0.83, 0.66 and 0.56, respectively. Compared with the segmentation performance of employing all four modalities in the axial view, it seems that there is no obvious advantage if not all modalities are adopted. \par

\begin{table*}[!htbp]
  \centering
  \caption{The comparison of other counterpart methods on the testing dataset in BRATS 2015 challenge.}
    \resizebox{0.99 \textwidth}{!}{ 
    \begin{tabular}{ccccccc}
    \toprule
    \multirow{2}[4]{*}{case} & \multirow{2}[4]{*}{method} & \multicolumn{3}{c}{dice} & \multirow{2}[4]{*}{segmentation time} & \multirow{2}[4]{*}{device} \\
\cmidrule{3-5}          &       & complete &  core  & enhancing &       &  \\
    \midrule
    \multirow{2}[2]{*}{53} & A Novel CNN-based Method \cite{ref15} & 0.78  & 0.65  & 0.75  & $<$20s & GPU \\
          & InputCascadeCNN \cite{ref14} & 0.79  & 0.58  & 0.69  & 25s-3min  & GPU \\
    \midrule
    \multirow{4}[2]{*}{110} & DeepMedic \cite{ref32} & 0.847  & 0.670  & 0.629  & $<$30s & GPU (at least 12 GB) \\
          & Ensemble \cite{ref32} & 0.849  & 0.667  & \textbf{0.634} & $<$30s & GPU (at least 12 GB) \\
          & FCNN(fusing) \cite{ref17} & 0.84  & \textbf{0.73} & 0.62  & 2-4min & GPU \\
          & Our method* & \textbf{0.85} & 0.71  & 0.61  & \textbf{$<$15s} & GPU (at least 8 GB) \\
    \bottomrule
    \end{tabular}
    }
  \label{tab:other_methods}
\end{table*}

Furthermore, in \cite{ref32,ref17}, five smaller image patches are truncated from the upper left, the lower left, the middle, the upper right and the lower right of the center area from the input image, and these patches are regarded as the supplementary input in the second stage of network downsampling. This smaller image patch method is also adopted in this work to improve the proposed multi-view framework. As seen from the second and third rows in Table \ref{tab:others_improvements}, it can be found that the smaller image patches method achieves at least 1\% promotion on the complete, core and enhancing regions when segmenting the brain tumor from the single view. Especially for the core tumor region, the segmentation performance is raised by 2\% on the dice metric. And the experimental results of adopting smaller image patches on the segmentation network from multi-views can be found in the fourth and fifth views. By comparing the segmentation performance with/without the smaller image patches, it can be found that the dice metric of the complete and core regions is improved by 1\% and the best segmentation performance is 0.85, 0.69 and 0.61, respectively. These experimental results prove that adopting the smaller image patches as the supplementary input is an effective improvement for the proposed multi-view segmentation framework. Moreover, the reason behind this is that the problem of unbalanced data categories can be solved in some ways by adopting the smaller image patches and the better segmentation performance can be achieved.  

Besides for the aforementioned improvement, the post-processing methods proposed by \cite{ref17} is also employed in this paper. Different from the original post-processing method, in this paper, only the 3rd, 4th and 6th steps are adopted to implement the post-processing. As seen from the last row in Table \ref{tab:others_improvements}, the dice metric of core tumor region is raised from the 0.69 to 0.71 after adopting the post-processing method. 

Overall, by considering all improvements for the proposed multi-view framework, the best segmentation performance can be achieved with the dice metric of 0.85, 0.71 and 0.61 on the complete, core, enhancing tumor regions, respectively.

\subsection{Comparison With Other Methods}

In the BRATS 2015 datasets, there are only 53 testing cases available at the beginning while this number is increased to the 110 at later. More testing cases result in a more challenging segmentation task. As seen in Table \ref{tab:other_methods}, the \cite{ref14} and \cite{ref15} conducted the experiments on the 53 cases while the \cite{ref32}, \cite{ref17} and our method were evaluated on the 110 cases. It can be found the proposed method achieves a better segmentation performance than the \cite{ref14} and \cite{ref15} even if faced with a more challenging task. Furthermore, compared to other two methods evaluated on the 110 cases, our method also achieves the competitive advantage in the segmentation performance, especially our method obtains the highest value 0.85 on the dice metric of complete tumor region. And the dice metrics of core and enhancing tumor regions are very close to the best one obtained by other methods. The reason behind this shortage may be caused by the preprocessing step with the N4ITK method adopted by \cite{ref17} and the post-processing step with 3D CRF method adopted by \cite{ref32} and \cite{ref17}. The preprocessing N4ITK method can enhance the image quality and the 3D CRF method has the ability to further improve the segmentation performance. However, these methods greatly increase the computational burden for brain tumor segmentation and are also time-consuming task. As a contrast, the proposed method just adopts some simple yet effective preprocessing and post-processing method to reduce the computational burden and improve the segmentation efficiency. As shown in Table \ref{tab:other_methods}, it can be seen that the proposed method only costs less than 15 seconds to segment an entire brain, which is more than twice as fast as other counterpart methods. It means that the proposed method also achieves a high performance in term of efficiency. Overall, considering both the effectiveness and the efficiency, the multi-view dynamic fusion framework is a better choice for brain tumor segmentation compared to other state-of-the-art counterpart method. \par

\begin{table}[htbp]
  \centering
  \caption{The segmentation performance of the proposed method on the BRATS 2018.}
    \begin{tabular}{ccccc}
    \toprule
    \multirow{2}[4]{*}{dataset} & \multirow{2}[4]{*}{case} & \multicolumn{3}{c}{dice} \\
\cmidrule{3-5}          &       & complete &  core  & enhancing \\
    \midrule
    BRATS 2015 & 110   & \textbf{0.85}  & 0.71  & 0.61  \\
    BRATS 2018(UNET axial view) & 66    & 0.79  & 0.69   & 0.65 \\
    BRATS 2018(Our method) & 66    & 0.84  & \textbf{0.74}   & \textbf{0.71} \\
    \bottomrule
    \end{tabular}
  \label{tab:brats2018}
\end{table}

Besides the BRATS 2015, the proposed method is also evaluated on the BRATS 2018 with the online validation dataset. As shown in Table \ref{tab:brats2018}, it can be found that the proposed method also obtains a good performance on the BRATS 2018, which proves the generalization of our method.

\section{Conclusion} \label{section:Conclusion}

Inspired by the way that humans observe space objects and clinicians diagnose the brain from multi-view, we discussed how to adopt the multi-views method to improve the performance of brain tumor segmentation. The multi-view dynamic decision fusion framework has been proposed to facilitate the process of implementing the multi-view method. The proposed framework can be used to dynamically fuse segmentation results obtained from 2D brain tumor images corresponding to different views, such as the axial view. Besides dynamically fusing segmentation results from multi-views, the improved fusion loss has also been employed to train network so as to keep the appearance and space consistency on the training process. \par

According to the experiments on BRATS 2015, it can be seen that for the segmentation result obtained from the single view, the best performance can be achieved in the axial view, followed by the coronal view, and the worst is the sagittal view. After adopting brain images from multi-views, the integrated results from three views are superior to any one from the single view or from just two views. In other words, the multi-view method has the ability to improve the brain tumor segmentation. In addition, the experiments also confirmed that the proposed transition loss and the decision loss were a benefit to brain tumor segmentation with multi-view methods. \par
Besides the improvement in the framework, some simple yet effective methods have been evaluated so as to achieve a better performance. In addition, the proposed method performed well in term of efficiency. Overall, it can be proven that the multi-view dynamic fusion framework is a good choice for segmenting the brain tumor compared to other counterpart methods. Moreover, the proposed framework is also evaluated on the BRATS 2018. According to the experiment on the on-line validation dataset, it can be proven that the proposed framework also achieves a good performance for segmenting the brain tumor.\par
In the future, our ongoing research will focus on how to further take advantage of features extracted from multi-views and how to reduce the computing parameters of the network model. 

\section{Acknowledgements}
This work was supported by the Natural Science Foundation of Guangdong Province (Grant No.2018A030313354), the Neijiang Intelligent Showmanship Service Platform Project (No.180589), the Sichuan Science-Technology Support Plan Program (No.2019YJ0636, No.2018GZ0236, No.18ZDYF2558), and the National Science Foundation of China - Guangdong Joint Foundation (No.U1401257).

\bibliographystyle{unsrt}  
\bibliography{references}


\end{document}